# Quantum Simulations of SARS-CoV-2 Main Protease M^pro Enable Accurate Scoring of Diverse Ligands

Yuhang Wang[1], Sruthi Murlidaran[1], and David A. Pearlman[1*]

[1]*Quantum Simulation Technologies, Inc., 625 Massachusetts Ave, Floor 2, Cambridge, MA 02139*

A B S T R A C T

The COVID-19 pandemic has led to unprecedented efforts to identify drugs that can reduce its associated morbidity/mortality rate. Computational chemistry approaches hold the potential for triaging potential candidates far more quickly than their experimental counterparts. These methods have been widely used to search for small molecules that can inhibit critical proteins involved in the SARS-CoV-2 replication cycle. An important target is the SARS-CoV-2 main protease M^pro, an enzyme that cleaves the viral polyproteins into individual proteins required for viral replication and transcription. Unfortunately, standard computational screening methods face difficulties in ranking diverse ligands to a receptor due to disparate ligand scaffolds and varying charge states. Here, we describe full density functional quantum mechanical (DFT/QM) simulations of M^pro in complex with various ligands to obtain absolute ligand binding energies. Our calculations are enabled by a new cloud-native parallel DFT/QM implementation running on computational resources from Amazon Web Services (AWS). The results we obtain are promising: the approach is quite capable of scoring a very diverse set of existing drug compounds for their affinities to M^pro and suggest the DFT/QM approach is potentially more broadly applicable to repurpose screening against this target. In addition, each DFT/QM simulation required only ~1 hour (wall clock time) per ligand. The fast turnaround time raises the practical possibility of a broad application of large-scale quantum mechanics in the drug discovery pipeline at stages where ligand diversity is essential.

## 1. Introduction

Computational chemistry has made significant progress in the past several decades, addressing bottlenecks in the drug discovery process. The improvement is particularly visible in the ligand triage step during the initial virtual screening phases (e.g., via molecular docking). The increased use of computational chemistry techniques is also seen in binary decision making near the end of a drug discovery project when the molecular scaffold has been established, and one seeks only to compare congeneric compounds (e.g., via free energy calculations). However, there is a significant computational gap in the middle of the discovery process, where more diverse compounds are encountered. There is an urgent need for computational approaches that can rank order dozens, or hundreds, of *unrelated* compounds, with sufficiently high accuracy[1].

The technical requirements of such an approach are that it should be able to (1) score ligands with diverse scaffolds; (2) deal with variances in formal charge and polarization; (3) be applicable to realistic models of ligand/protein interactions; and (4) perform these calculations sufficiently quickly to be compatible with modern drug discovery, all while retaining good accuracy. Existing, widely used methods, such as those based on free energy perturbation and classical force fields[2,3], usually satisfy criteria 3-4 but fail 1-2. In principle, high-level quantum mechanical calculations, at the level of modern density functional theory (DFT/QM), can address both 1-2, but, until recently, could not be performed on large enough systems with sufficient throughput to address points 3-4[4,5]. In a recent publication[6], we described an implementation of a new algorithm for quantum calculations ("high-efficiency distributed QM" (hedQM)). This implementation allows quantum determinations at the DFT/QM level to be performed with reasonable throughput (~1 hour) on systems much larger than ever before possible - for example, full proteins - enabled by easily accessible commercial cloud compute resources, such as those offered by Amazon Web Services (AWS).

Here, we apply this method to a data set germane to identifying new drugs that might help battle the COVID-19 virus. The dataset originates from a recent publication[7], and we briefly describe its construction here. A set of more than 2,500 drug molecules previously approved for various applications were subjected to a computational screen against $M^{pro}$, the SARS-CoV-2 main protease[8]. This enzyme cleaves viral polyproteins into individual proteins required for viral replications and transcription, and it is hypothesized that inhibiting this protein would inhibit replication of the COVID-19 virus[9]. From this computational screen, 100 molecules were identified as having the potential to bind to $M^{pro}$ using a combination of molecular docking and absolute binding free energy calculations. Subsequently, this set of ligands was screened experimentally, leading to a set of 16 molecules with measurable binding to the $M^{pro}$ receptor. With experimentally determined binding affinity values for $M^{pro}$, this set of ligands serves as the validation set for this study. The chemical structures of these drug molecules are shown in Figure 1.

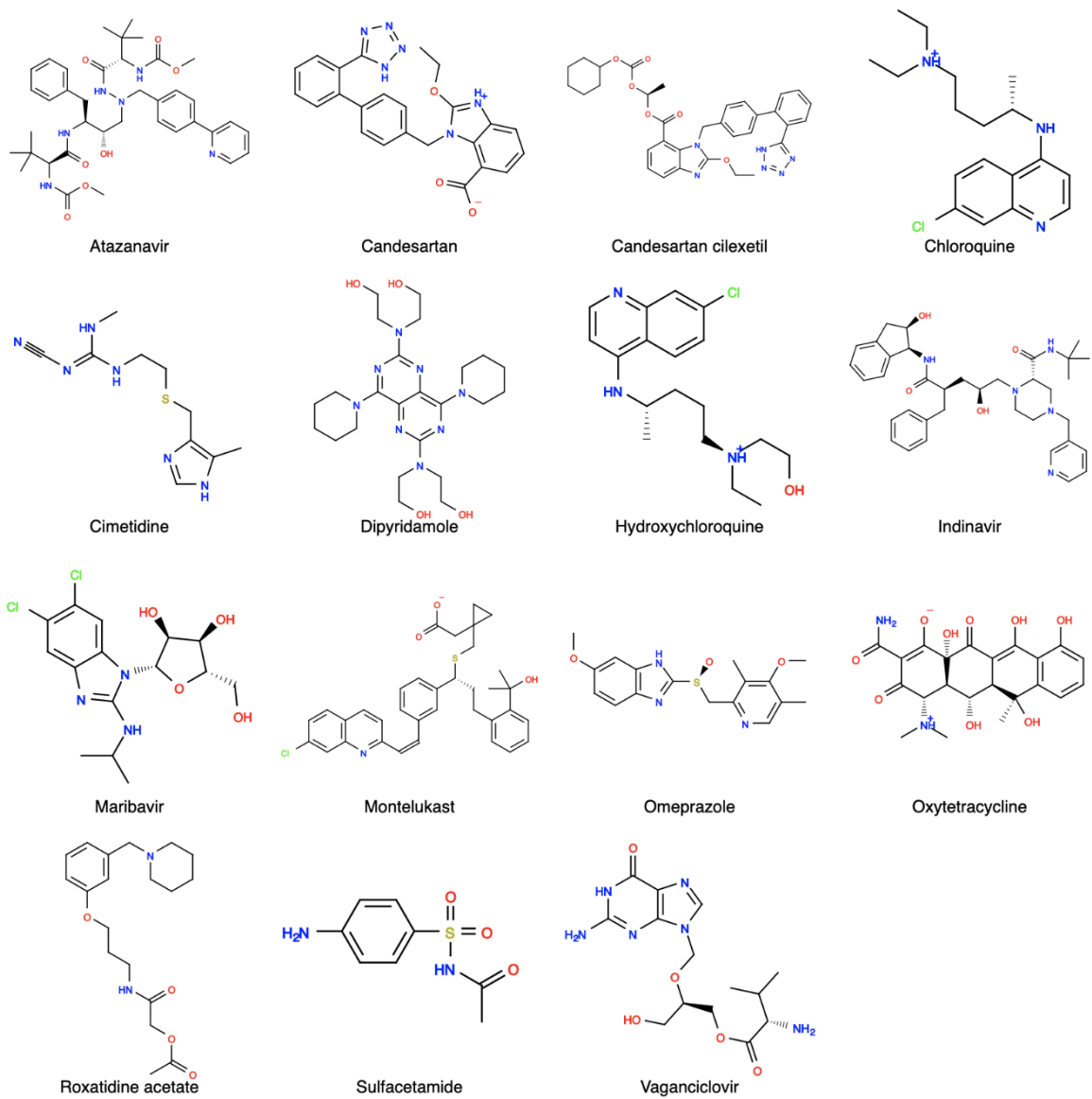

**Figure 1**. Drug molecules examined in this study. Each has been experimentally determined[7] to bind with measurable affinity to SARS-CoV-2 $M^{pro}$. For each ligand, several isomers were considered, leading to a range of charge states.

Two observations can immediately be made about this set of molecules. First, they are extremely diverse and reflect a highly divergent set of scaffold classes. Second, they also reflect a diversity of charge states across the potential isomers. These challenges are precisely those raised in points 1-2 above. In particular, the diversity of charge states makes it an extremely challenging data set to model by standard force-fields and associated molecular mechanics methods[10]. The multiple scaffold classes render the set difficult or impossible to address with relative difference methods like FEP, which typically require that the ligands being studied be fairly similar to one another[11]. For these reasons, this is a data set that has the potential to demonstrate the value that a high-level quantum approach –-capable of determining absolute energies of binding -- can bring to such an investigation.

## 2. Methods

The set of 16 drug compounds with experimentally measured binding to $M^{pro}$ is taken from a recent study[7], as are the experimental ligand binding free energies. One compound (disulfiram) reported in that publication is omitted because it is believed to be covalently bound[12]. For the remaining 15 compounds, we applied our cloud-native parallel hedQM approach to determine the absolute energy of binding at the DFT/QM level. We used the revPBE functional[13] with the D3(BJ) dispersion correction[14], the def2-SVP basis[15] in a 9.1 Å sphere around the ligand within the binding site, and a minimal basis (MINAO)[16] for atoms outside this sphere. Details of the DFT/QM calculations are provided in the supplemental information.

Since experimental structures of the bound ligand/protein complexes were not available, it was necessary to generate them using structure-based docking. A crystallographic structure of the $M^{pro}$ protein (Mpro-x3080) was obtained from the Diamond Light Source (UK) synchrotron facility's Fragalysis web application[17]. These crystal structures of $M^{pro}$ were determined as part of the COVID Moonshot project[18]. AutoDock Bias method[19] was then used for docking. The $M^{pro}$ protein structure consists of a domain including the binding site and a second alpha-helical domain located far from the binding site. In solution, the protein forms a homodimer. To optimize computational cost, we truncated the beginning of the unstructured N-terminal region (SER1 to LYS5) and the alpha-helical domain (residue ASP197 to THR304). The new terminal residues (MET6 and THR196) were capped with ACE and NME terminal patches, respectively. The protein structure is shown in Figure 2. The truncated protein, which retains the active site, contains 2900 atoms. The atoms shown in gray are those truncated off for the calculations.

For each ligand, we included several isomers, and each isomer was processed independently during the docking process, then all poses for the same parent molecule were aggregated for the subsequent scoring process. A total of 100 docked complexes were generated for each ligand. To rank the docked poses, we first evaluated the total energies of the docked structures at the molecular mechanics level. Then we selected the top-50 docked poses and tightly minimized the structures using molecular mechanics (see supplemental information). From the resulting set of 50 MM-minimized docked structures, the ten lowest energy ligand/protein poses were further optimized using the semi-empirical GFN1-xTB method[20]. The two lowest-energy GFN1-xTB ligand/protein poses for each ligand were then selected for full DFT/QM calculation. The post-docking classical mechanics calculations were carried out using AmberTools20[21], using the Generalized Born implicit solvent model[22] (igb=5), the Amber 14 force field[23], and the GAFF force field[24] for ligands, assigned using Antechamber from AmberTools.

To determine the lowest energy conformation of the unbound ligand, we used a combination of conformers generated from the classical mechanics search method RDKit[25] and the semi-empirical conformational search protocol in CREST[27]. From the set of resultant conformers, the energies of the ten lowest energy structures were recalculated using DFT/QM using the C-PCM implicit solvent model[28].

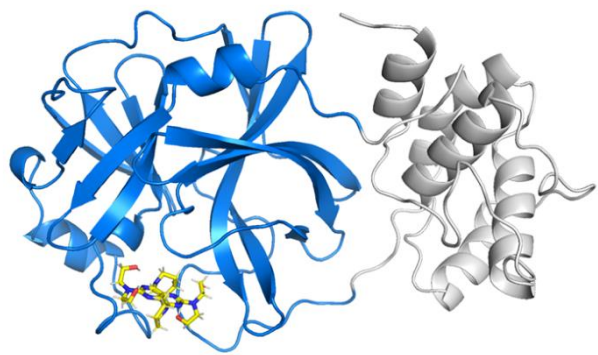

**Figure 2**. $M^{pro}$ protein with a ligand (dipyridamole) bound to its active site. The region in gray was excluded from all calculations. The ligand is shown in licorice representation (image generated using PyMOL[26]).

The net energy of binding is determined from the relationship:

$$\Delta E \ (P+L \rightarrow P \cdot L) = E(P \cdot L) - E(P_{complex}) - E(L_{min}) \qquad (1)$$

where $E(P \cdot L)$ is the energy of the complex, $E(P_{complex})$ is the energy of the protein alone, in the same conformation as the complex, and $E(L_{min})$ is the minimum energy of unbound ligand conformer, determined using the search approach described above. To reflect conformational sampling, we used Boltzmann averaging for the two most favorable docked poses for each ligand isomer, and linear averaging for the ligand isomers (this avoids difficulties in reweighting energies of isomers with different numbers of atoms).

## 3. Results

The net binding energies calculated using full DFT/QM are presented in Figure 3. We use this as a ranking score and plot it against the experimental binding free energies in a correlation plot. For comparison, the same plot is presented for the semi-empirical GFN1-xTB quantum method in Figure 4.

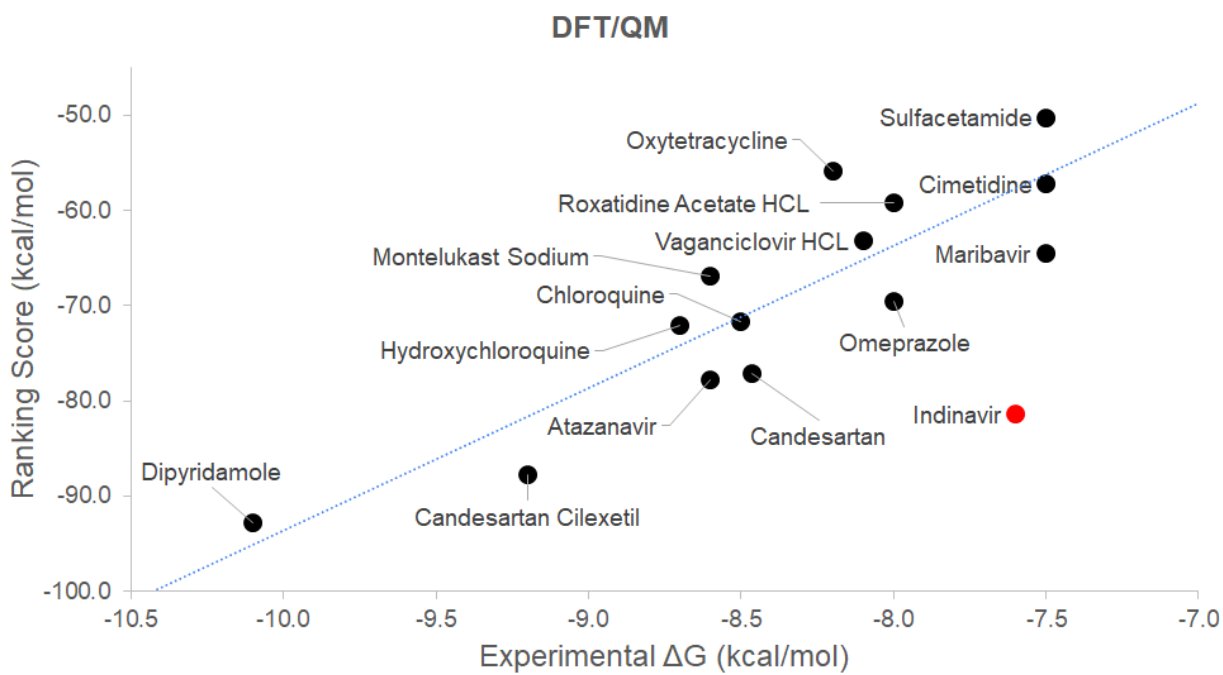

**Figure 3**: Binding energies predicted using DFT/QM. The trend line is calculated excluding the outlier indinavir. The overall $R^2$ value for all points (including indinavir) is 0.55. The $R^2$ value excluding indinavir is 0.77. The Predictive Indices with and without indinavir are 0.74 and 0.88, respectively.

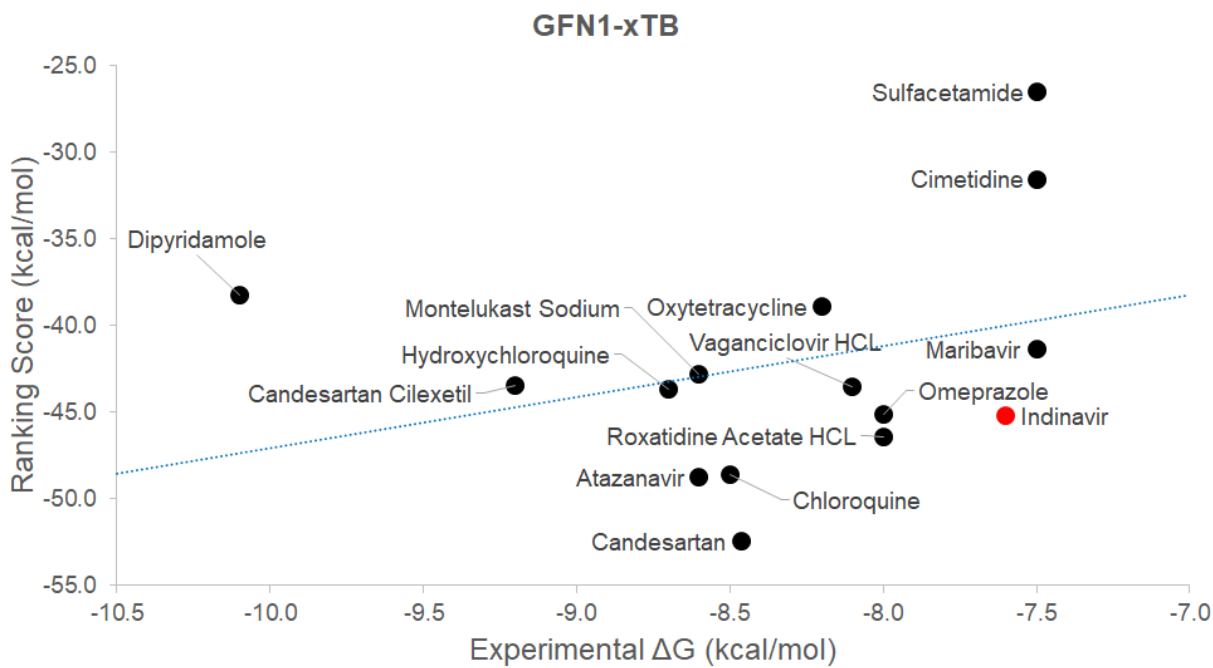

**Figure 4**: Binding energies predicted using GFN1-xTB. The trend line is plotted is calculated excluding the outlier indinavir. Overall $R^2$ value for all points (including indinavir) is 0.07. The $R^2$ value excluding indinavir is 0.09. The Predictive Indices with and without indinavir are 0.08 and 0.15, respectively.

As can be seen, the correlation obtained using DFT/QM calculations of the COVID binding domain is very good, with an $R^2$ value of 0.55, and a Predictive Index[29] (a weighted measure of the ability of a predictor to properly rank order) of 0.74. Only one ligand (indinavir) falls far off the correlation line for reasons that are not clear. It is particularly satisfying to observe that DFT/QM very clearly differentiates the two ligands that bind best experimentally (dipyridamole and candesartan cilexetil) from the remaining ligands. In contrast, the simpler semi-empirical QM approximation fails to capture the correlation for this ligand set, with an R2 of 0.07 and a Predictive Index of 0.08.

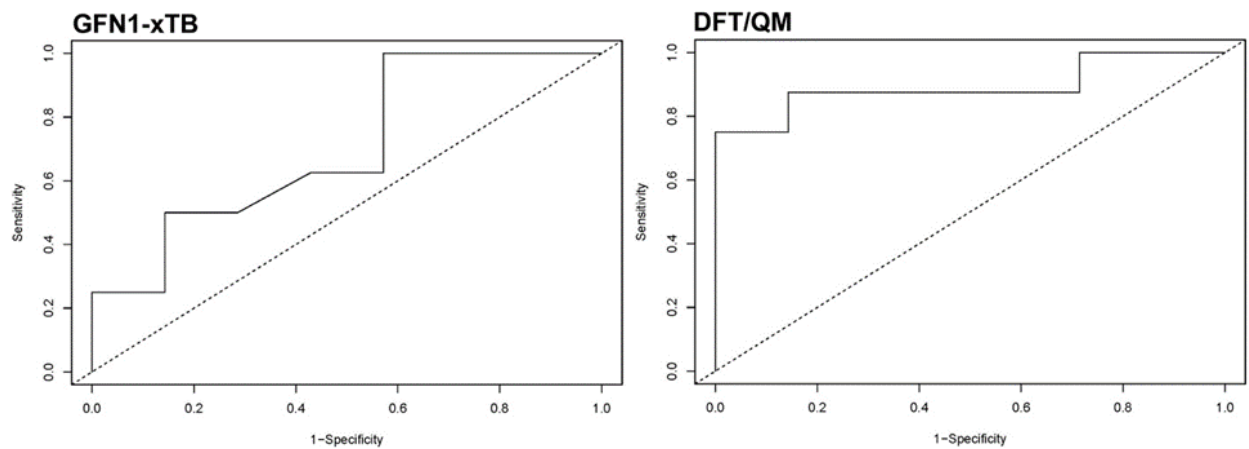

**Figure 5**. The computed ROC curves for the GFN1-xTB and DFT/QM data are shown in Figure 4, with 7 ligands that experimentally best bind (Ki < 800 nM) designated as true positives. The AUC for GFN1-xTB (left) is 0.71, while the AUC for DFT/QM (right) is 0.89.

Evaluation of the DFT/QM results using ROC analysis[30] further corroborates this analysis. If we take the best half of the binders (top 7) and designate them as the "hits" (equivalent to designating all binders that bind better than 800nM as hits), we get the curves shown in Figure 5. The area under the curve for DFT/QM is 0.89, reflecting an excellent ability to differentiate the better binders from the remaining ligands in this set.

In addition to the QM-based approaches we have applied, the publication that identified this data set[7] described the application of a classical mechanics FEP-based approach to the determination of absolute binding free energies for this set. They obtained some signal with their approach, with an $R^2$ of 0.31 and a Predictive Index of 0.66, although the results we obtained with DFT/QM significantly improve on this. Given the diversity in formal charges among the ligands, which is better addressed with QM, this is not surprising.

To further understand the origins of the differing predictions using semi-empirical GFN1-xTB and full DFT/QM, in Fig. 6 we plot the binding energies computed using the two methods against each other. We color code the ligands by their charges, and we plot the energies of the different isomers considered for each ligand separately (as they may have different charges). Points in the plot are colored black for neutral, orange for positively charged, and green for negatively charged. As can be seen from the plot, the correlation between the semi-empirical and DFT/QM binding energies is moderate when limited to either the neutral or positively charged ligands but is very poor when multiple charge types are considered. For the negatively charged ligands, the correlation is even worse. This is consistent with the fact that charged systems, and especially anions, are more challenging for semi-empirical methods because the corresponding wavefunctions and charge densities are parametrized by atomic minimal bases that cannot fully respond to large polarization effects.

As noted earlier, comparing multiple ligands with varying net charges is well-known to be challenging in comparative analysis. It is thus reassuring to observe that the charge of the ligand does not bias the DFT/QM-based binding affinity predictions. This is, of course, a fundamental advantage of QM when compared to force-fields, and of full DFT/QM calculations that use realistic bases when compared to semi-empirical quantum approximations. The predictive power of DFT/QM on this set is particularly noteworthy because of the uncertainties associated with the conformations for both the bound ligand/complex and the unbound ligand and because the energy determinations are single-point energies, effectively at 0° K with no entropic contribution

*\* Corresponding author.*
E-mail address: pearlman@qsimulate.com

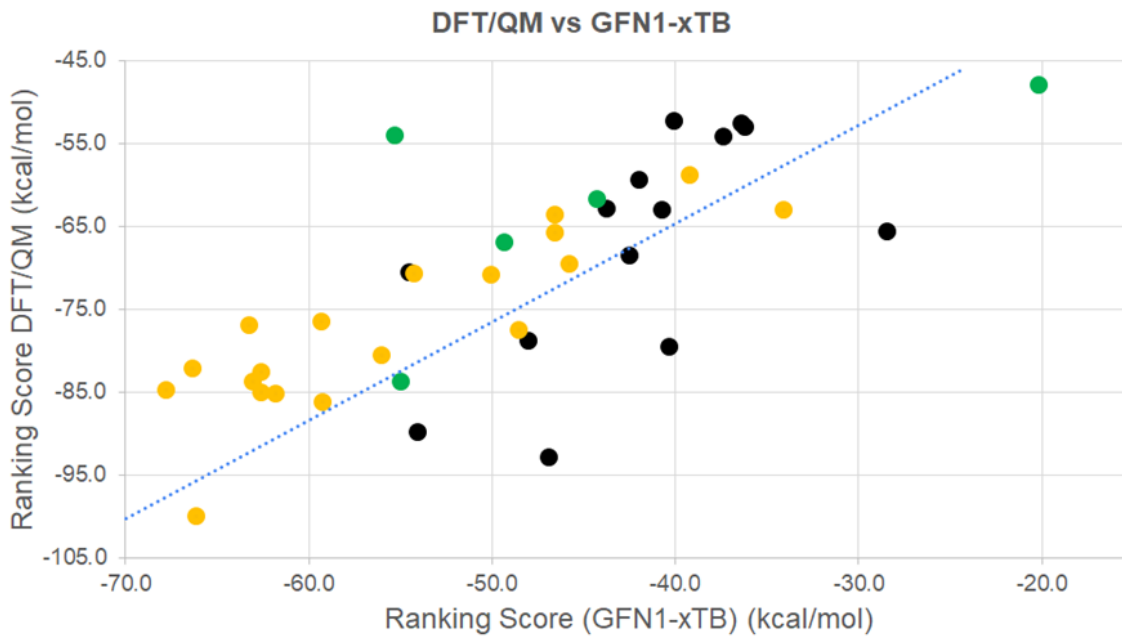

**Figure 6**: A comparison of the scores calculated using the semi-empirical approach (GFN1-xTB) and DFT/QM. Many of the 15 ligands are represented by multiple data points, corresponding to multiple isomers of that ligand. Black: neutral ligand isomers. Orange: positively charged ligand isomers. Green: negatively charged ligand isomers.

## 4. Discussion

We have demonstrated that high-level quantum mechanics (density functional theory with dispersion corrections, using a realistic basis) can be successfully applied to rank order a scaffold-diverse set of ligands to a realistic protein receptor model—focusing on a set of existing drugs that are known to bind to the COVID-19 relevant protein $M^{pro}$. The full density functional treatment provides results that are substantially better than those obtained using a semi-empirical quantum approximation (which show almost no correlation for this dataset). Despite a quantum mechanical domain of nearly 3000 atoms, these calculations were carried out with realistic turnaround times and modest, accessible cloud-based computational resources using our recently described parallel implementation of DFT/QM quantum mechanics. Each ligand isomer calculation required ~1 hour in wall clock time with 14 AWS r5.24xlarge instances, with a compute cost of less than $90 (On-Demand instances) or $15 (Spot Instances).

In light of this performance, scope of applicability, and throughput, one could envision running a fully quantum-based screening campaign on hundreds, or even thousands, of compounds, with diverse scaffolds, charges, and chemical structures—an endeavour that would be extremely difficult or impossible using current methods based on force-fields. The flexibility of the quantum mechanical approach thus offers potentially new ways to use computation to advance drug discovery.

The idea that QM can be applied to drug discovery is not a new one. But earlier efforts have had to make a variety of compromises, e.g., via semi-empirical energy functions, fragment or linear-scaling approximations that introduce substantial cutoffs[31] or else have restricted QM to a small nucleus in QM/MM treatments[5]. In addition, the observed turnaround time using these approaches has typically been unrealistically long, on the timescale of days or weeks. These compromises have been an obstacle to realizing the predictive potential of QM in the context of drug discovery. For example, we see that semi-empirical parameterizations, even in their modern incarnations such as GFN1-xTB, lead to substantial errors in evaluating interactions such as charge transfer that are required to suitably assess diverse ligands. Similarly, QM/MM or fragmentation methods introduce errors caused by artificial boundaries and inaccurate treatment of long-range charge polarization[31-33]. What we have now demonstrated is that it is practical to treat a substantial region of a ligand/protein system – several thousand atoms - with full DFT/QM without introducing compromises and on a realistic computational timescale, a significant practical advance over previous applications of QM to drug discovery.

\* *Corresponding author.*
E-mail address: pearlman@qsimulate.com

It is important to note that although the DFT/QM calculations described herein performed quite well, these calculations only predict the 0º K enthalpy of binding. Entropic contributions, including desolvation of the binding pocket, as well as the entropic changes arising from conformational variability of the ligand and protein, have not been included. The quality of results for this set suggests that these contributions may be of minor importance for this protein target and ligand set, but looking more broadly, there will assuredly be systems where that is not the case. To address these issues, we are currently working on integrating corrections to the approach we have used to account for the desolvation entropy and for the entropic contributions of the ligand and protein. We will report on these improvements in a future publication.

## 5. Acknowledgements

We thank Garnet Chan for his continual and extensive advice and discussion as this work was carried out and Toru Shiozaki for his assistance with the quantum calculations and his constant encouragement of this work. We also thank Satish Gandhi at Amazon Web Services (AWS), Simone Severini, Eric Kessler, and other members of the AWS Quantum Technologies team, and the AWS organization for the very generous computational support that made this work possible.